# Proteome remodelling during development from blood to insect-form *Trypanosoma brucei* quantified by SILAC and mass spectrometry

Kapila Gunasekera[1,3], Daniel Wüthrich[1], Sophie Braga-Lagache[2], Manfred Heller[2] and Torsten Ochsenreiter[1*]

## Abstract

**Background:** *Trypanosoma brucei* is the causative agent of human African sleeping sickness and Nagana in cattle. In addition to being an important pathogen *T. brucei* has developed into a model system in cell biology.

**Results:** Using Stable Isotope Labelling of Amino acids in Cell culture (SILAC) in combination with mass spectrometry we determined the abundance of >1600 proteins in the long slender (LS), short stumpy (SS) mammalian bloodstream form stages relative to the procyclic (PC) insect-form stage. In total we identified 2645 proteins, corresponding to ~30% of the total proteome and for the first time present a comprehensive overview of relative protein levels in three life stages of the parasite.

**Conclusions:** We can show the extent of pre-adaptation in the SS cells, especially at the level of the mitochondrial proteome. The comparison to a previously published report on monomorphic *in vitro* grown bloodstream and procyclic *T. brucei* indicates a loss of stringent regulation particularly of mitochondrial proteins in these cells when compared to the pleomorphic *in vivo* situation. In order to better understand the different levels of gene expression regulation in this organism we compared mRNA steady state abundance with the relative protein abundance-changes and detected moderate but significant correlation indicating that trypanosomes possess a significant repertoire of translational and posttranslational mechanisms to regulate protein abundance.

## Background

*T. brucei sspp* are single celled, flagellated protozoan parasites that cause human African sleeping sickness and Nagana in cattle. The organisms follow a complex life cycle alternating between the mammalian and insect hosts. *T. brucei sspp* replicate by binary fission and, in the infected mammals, long-slender bloodstream forms use antigenic variation of their surface protein coat to effectively evade elimination of the population by the host immune system. In order to efficiently infect the insect vector (tsetse; Glossina sp.) the LS cells (pleomorphic cell lines) transform into quiescent "short stumpy" (SS) cells that are pre-adapted to the life in the midgut of the tsetse. This transition occurs in response to a parasite derived unidentified "stumpy inducing factor" (SIF) in a cell density dependent manner [1]. The further transition from the quiescent SS stage to the procyclic (PC) insect stage occurs in the tsetse midgut. *In vitro* this situation can be mimicked through a temperature drop (from 37°C to 27°C) and citrate or cis-aconitate addition to the medium. The mechanism of the in vitro differentiation involves the stage-specific expression of carboxylate transport proteins [2] and a phosphatase cascade that eventually relays the differentiation signal to the glycosomes [3]. Glycosomes are peroxisome-derived trypanosome-specific organelles harbouring more than 200 proteins involved in a number of different pathways including Glycolysis and the beta-oxidation of fatty acids [4,5]. How the glycosomes further promote differentiation remains elusive.

During the life cycle *T. brucei* undergoes dramatic changes including the surface proteome, changes in overall size, shape and motility as well as intracellular changes most prominently seen in the mitochondrion. The organelle transforms from an acristate tubular structure devoid of cytochromes, oxidative phosphorylation and tricarboxylic acid "cycle" (TCA) activity to a

* Correspondence: torsten.ochsenreiter@izb.unibe.ch
[1]Institute for Cell Biology, University of Bern, Bern, Switzerland
Full list of author information is available at the end of the article compartment capable of the *bona fide* mitochondrial activities [6]. The morphological and metabolic changes occurring during the life cycle are accompanied by major changes in gene expression at the mRNA level [7-12]. Trypanosomes regulate steady state mRNA abundance mainly at the posttranscriptional level likely through differential stability of the mRNA molecules as has been shown for a number of transcripts [13] [14,15]. Additionally, a number of individual studies including work on the procyclic surface proteins [16], the glycosomal aconitase [17], the mitochondrial cytochrome c reductase [18] and several cytochrome oxidase subunits [19], have demonstrated that trypanosomes use translational and posttranslational mechanisms to regulate gene expression, however the genome-wide scale of these mechanisms remains unknown. The majority of proteomic studies in *T. brucei* have focused on individual parts/compartments of the cell including the mitochondrion [20], the plasma membrane [21], the flagellum [22,23] or applied semi-quantitative strategies in two life stages (LS and PC; [24,25]). No comprehensive proteomics study of the three life cycle stages have been done and/or compared with the corresponding transcriptome. Thus the importance and scale of translational regulation and protein stability as means of gene expression regulation in *T. brucei* remain elusive.

In recent years SILAC approaches have become the method of choice for the quantitation of proteomes in numerous organisms ranging from yeast to plants and humans [26-28]. Here we present the first quantitative proteomics study of three life stages of *T. brucei* using SILAC and compare it with recently published transcriptome data and proteome data.

## Results

Insect-form *T. brucei* cells grew at very similar rates irrespective of the medium (SDM80 or 79), dialyzed/regular fetal bovine serum (FCS) and labeled or unlabeled amino acids (Additional file 1: Figure S1). We could show that after 11 cell division cycles in medium containing heavy isotope labeled Arg/Lys the mean ratio of heavy over light isotopes in the detected peptides changed more than 20 fold from 0.51 to 11.4 indicating that the majority of proteins had incorporated the heavy isotope amino acids (Additional file 1: Figure S2A). This is also supported by the inspection of individual peptides, where the incorporation of heavy isotopes can be followed from 0, 2, 11 cell cycle divisions (Additional file 1: Figure S2B). For the comparison of relative protein abundance we used *in vitro* grown PC (Antat 1.1) and the mammalian LS or SS parasites (pleomorphic Antat 1.1) from infected animals (Wistar rats). LS parasites were harvested from an early day three infection at a density of $5 \times 10^7$ cells/ml, while the SS cells were harvested at a density of $3\text{-}5 \times 10^8$/ml on day six from immunosuppressed animals. Parasites grew at similar rates in immunosuppressed and untreated animals (data not shown). We purified the bloodstream form trypanosomes using anion exchange chromatography as described previously [29]. $5 \times 10^6$ insect-form and bloodstream form cells (LS or SS) were lysed using SDS PAGE buffer and subsequently mixed. The total cell extracts were separated on SDS PAGE in triplicate. Each lane was cut into 10 slices, trypsinized and the resulting peptides analyzed by LC mass spectrometry.

### Overall proteome

To test how well the gel-base mass spectrometry approach covered the potential *T. brucei* proteome we compared the distribution of protein size, number of transmembrane domains and pI between the predicted and the detected proteome from two technical replicates (Figure 1A-C). We identified annotated proteins ranging from 71 (8.2 kDa, Tb11.01.1790) to 8214 amino acids (928 kDa; Tb09.160.1200), however, a bias against the detection of very small proteins (<8 kDa) and proteins with more than one trans-membrane domain were observed (Figure 1). Also very basic proteins rich in Arg and Lys residues were detected less often than predicted from the genome encoded proteome. In general we achieved very good coverage of the overall proteome, which can also be seen in the genome-wide plot of protein abundance-changes during the development (Figure 1D). The peptide information is summarized in the Additional file 2: Table S13. We detected uniform coverage along the 11 mega-chromosomes and the majority of transcription units, with the exception of the subtelomeric regions in most chromosomes.

In order to validate the mass spectrometry data and evaluate the whole-cell protein samples for integrity we compared the abundance of the proteins in mass spectrometry with the abundance in western blotting using a selected number of antibodies and found good correlation between the fold-change measured by the two techniques (Figure 2, Additional file 3: Table S1).

From the >2000 proteins detected (Additional file 4: Table S2 and Table S3) we reliably measured the abundance of 1673 in all three life stages (Additional file 5: Table S4). From these about 750 (44%) proteins were changed in expression level (≥ 2× up/down) in each of the comparisons, while about 900 (56%) proteins remained unchanged in abundance (Table 1). Among the unchanged protein groups we detected 102 ribosomal proteins with mean and median fold-change of close to one in each of the life-stage comparisons (Figure 3B and Additional file 6: Table S5). Analysis of the three life cycle stages showed a major change in protein composition during the transition from the proliferative LS to the SS cells (Figure 3A). More than 500 proteins were increased in abundance in the SS when compared to the LS, while only 192 were less abundant. The bias towards an increase in protein abundance was subsequently reversed during the transition from the quiescent SS to the insect-form (Figure 3A). A set of 650 proteins was detected only between LS and PC (Additional file 7: Table S6) and further 322 genes were detected only between SS and PC (Additional file 8: Table S7). The changes of protein abundance can also be grouped into the different cellular functions/compartments (GO terms; Figure 3B and Additional file 6: Table S5). Over 40% of the proteins in seven cellular component GO terms: plasma membrane, peroxisome, nucleus, mitochondrion, microtubule, lysosome and glycosome, are ≥ 2x increased in abundance during the differentiation to the short stumpy cells.

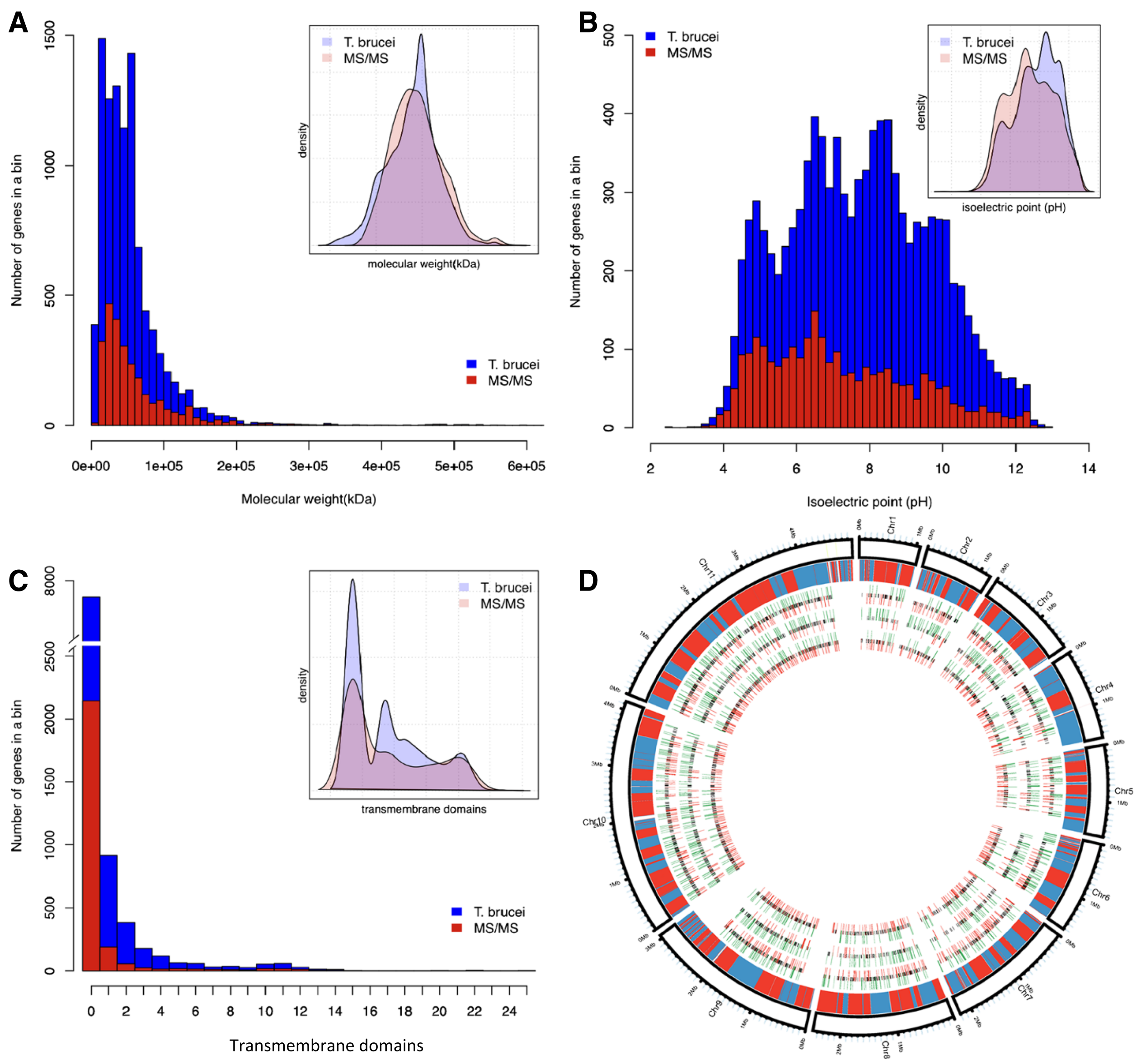

**Figure 1** ***T. brucei*** **proteome coverage using SILAC and mass spectrometry.** Comparison of mass spectrometry and genome encoded proteome. (**A**) molecular weight, (**B**) isoelectric point and (**C**) transmembrane domains. Histograms depict the number of proteins of a certain class encoded in the genome (blue) and detected by mass spectrometry (red). The insets depict a density plot of the corresponding histogram indicating which regions are over/under represented when comparing genome and mass spectrometry. (**D**) Distribution of protein ratios from the three life stages of *T. brucei* across the 11 mega-chromosomes. From outside to inside, the chromosomes one to 11, the polycistronic transcription units (PTUs) on the chromosomes (red, PTU on sense strand; blue, PTU on antisense strand), protein ratios LS/PC, SS/LS, PC/SS (green, ≥ 2x up; red, ≥ 2x down; black unchanged).

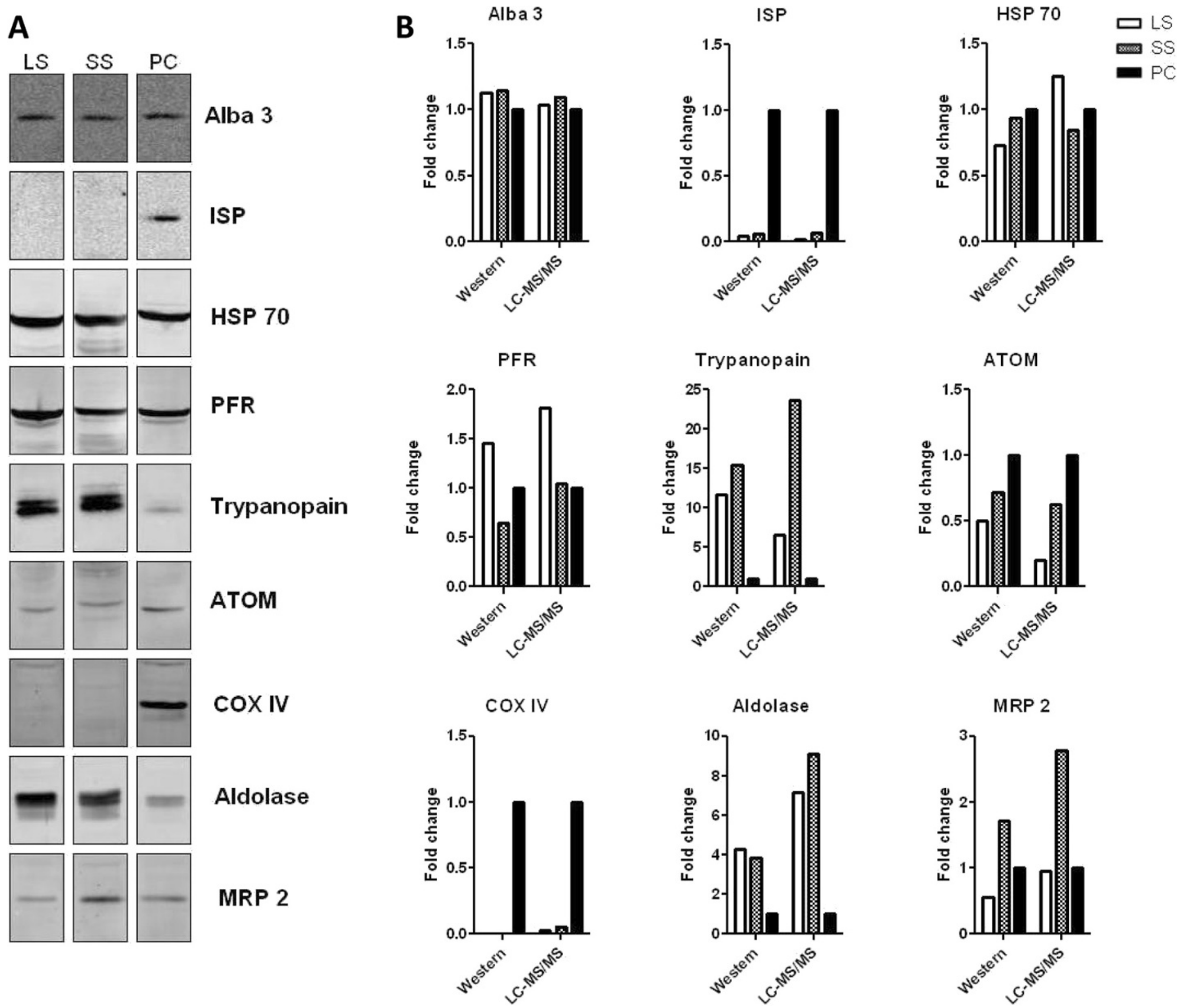

**Figure 2 Comparison of mass spectrometry and western blotting for protein abundance-changes during development of *T. brucei*.** Protein extracts used for mass spectrometry were probed using nine different antibodies. (**A**) Western blots using total cell extracts of long slender (LS), short stumpy (SS) and procyclic (PC). Antibodies used: Alba 3, iron sulfur protein (ISP), heat shock protein 70 (HSP 70), paraflagellar rod protein (PFR), trypanopain, archaic translocase of the outer mitochondrial membrane (ATOM), cytochrome oxidase subunit IV (COX IV), Aldolase and mitochondrial RNA binding protein 2 (MRP2). (**B**) Histograms of the fold-change measured by western blotting and mass spectrometry.

**Table 1 Overall proteome changes during development**

| | MS/MS (proteins) | | MS/MS overlap (proteins) | | | RNA-Seq and MS/MS overlap (proteins) | | |
|---|---|---|---|---|---|---|---|---|
| | **LS/PC** | **PC/SS** | **LS/PC** | **SS/LS[1]** | **PC/SS** | **LS/PC** | **SS/LS** | **PC/SS** |
| Detected | 2323 | 1995 | 1673 | 1673 | 1673 | 1529 | 1529 | 1529 |
| Up >2x | 564 | 308 | 362 | 559 | 250 | 321 | 498 | 218 |
| Down >2x | 551 | 645 | 398 | 192 | 507 | 354 | 174 | 449 |
| Unchanged | 1208 | 1042 | 913 | 922 | 916 | 854 | 857 | 862 |

[1] calculated from LS/PC and SS/PC.

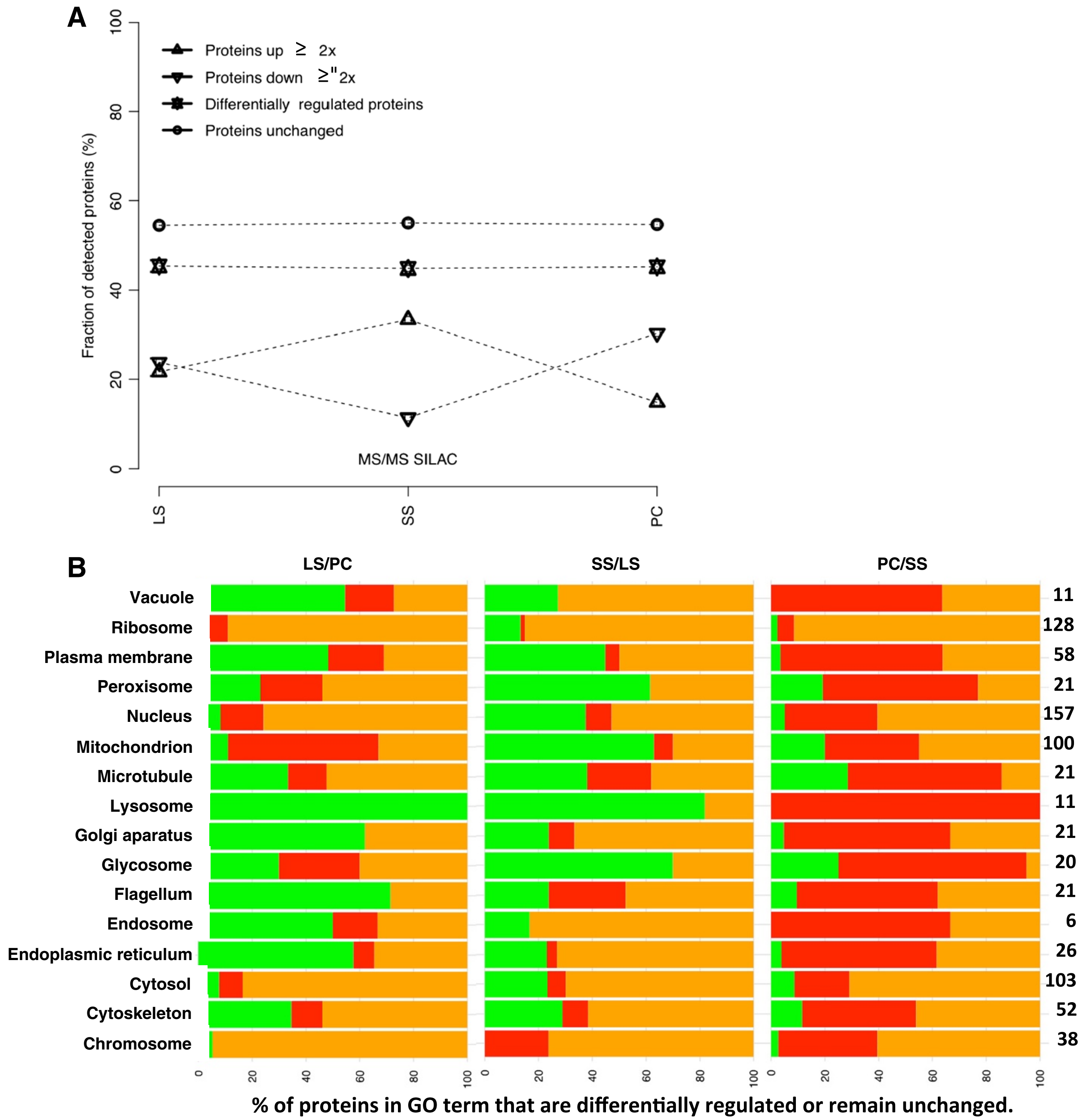

**Figure 3 Overall proteome changes during development and Gene Ontology (GO) term enrichment.** (**A**) Displayed are the fractions of proteins that change or remain unchanged in abundance during development. Additionally the direction of change is indicated (up/down). Figure corresponds to data shown in Table 1. (**B**) GO term enrichment. Green indicates fraction of proteins ≥ 2x upregulated, red (≥ 2x downregulated) and orange (unchanged).

We also compared our results to a recently published study by Urbaniak and co-workers [30], who applied a similar mass spectrometry approach to two life stages (LS and PC). Different from our study Urbaniak used *in vitro* grown monomorphic Mitat1.2 *T. brucei* cells that are not able to differentiate and complete the life cycle. Out of 2323 proteins that we detected in LS/PC 1851 could also be found in the study of Urbaniak and co-workers, while 1702 and 472 proteins were only detected in the Urbaniak and our study, respectively (Figure 4A). We compared abundance-changes of 1851 proteins and could show strong positive correlation for the vast majority of proteins between the two studies, excluding 22 proteins (1%) that displayed opposing directions of expression change (Overall Spearman's rank correlation: rho=0.83; Figure 4B). Both studies show an increase of mitochondrial protein abundance in the insect-form cells, however the increase is substantially larger for the study presented here (average of 8.5x) when compared to the study of the monomorphic cells (average of 3.5x). This can also be seen in Figure 4B) where 159 of 226 mitochondrial proteins detected (black circles) are above the diagonal that marks equal fold-change. 67 of 1851 proteins significantly differ in abundance between LS and PC in the two studies ($\Delta_{U/G}$PC/LS ≥ 5x; Additional file 9: Table S8). 50 and 17 of these proteins show larger fold-changes during the differentiation in the Gunasekera and Urbaniak study, respectively. 15 of these proteins (22%) are predicted to be mitochondrial. Additionally we compared relative abundance-changes during the transition from LS to PC from our study to the corresponding lifestages in *Trypanosoma congolense* from a recently published study [31] (Figure 4C-D). For 264 proteins the relative abundance-changes between LS and PC in the two parasite species correlated with a Spearman's rank correlation of rho=0.7 (p-value < 2.2e-16; Figure 4D), including 26 (10%) proteins that showed opposing patterns of expression in the two studies (Figure 4D). When we excluded those 26 proteins from the analysis the Spearman's rank correlation increased to rho= 0.87 (p-value < 2.2e-16).

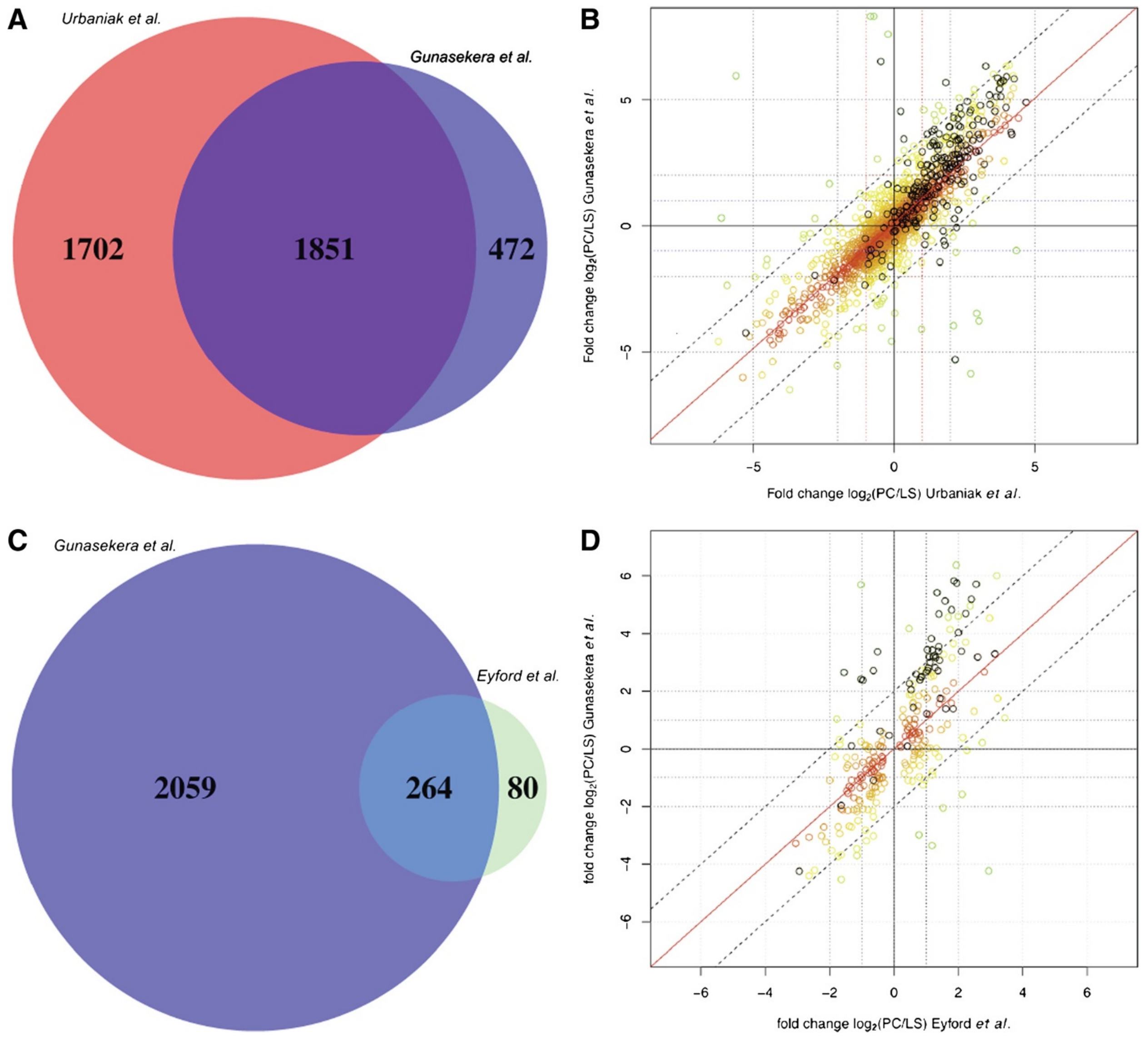

**Figure 4 Comparisons of the detection level and the protein abundance changes in three independent studies.** (**A**) Venn diagram displaying proteins detected in Urbaniak *et al*. and our study. (**B**) Correlation of protein abundance changes in the two studies. Colors of the circles in the scatter plot indicate correlation between the two studies (red high correlation, green low correlation). Black circles indicate mitochondrial proteins. The Spearman's rank correlation rho=0.82 (p-value < 2.2e-16). (**C**) Venn diagram displaying proteins detected in Eyford *et al*. and our study. (**D**) Correlation of protein abundance changes in the two studies. The Spearman's rank correlation rho=0.7 (p-value < 2.2e-16). For colors of the circles see (**B**).

### N-terminal extensions and alternative splicing to diversify protein production

Previously, others and we predicted several hundred proteins in the *T. brucei* proteome to have potential N-terminal extensions to the currently annotated start codon [7-9]. For 109 of these proteins we have detected N-terminal peptides covering the predicted extensions (Additional file 10: Table S9). Nineteen of the corresponding transcripts, including six MitoCarta transcripts contain alternative trans-splice sites downstream of the 5' most (distal) ATG (Figure 5). In the four representative examples alternative trans-splicing at the downstream positions would not permit translation of the N-terminal extension however, translation could initiate at one of the in-frame ATG codons downstream of the alternative splice site (Figure 5). One of the consequences of the change in the N-terminal peptide is exemplified with the gene Tb927.4.1410 that harbours a mitochondrial targeting signal in the N-terminal 45 amino acids. Translation from the alternatively trans-spliced transcript would not contain this targeting signal and thus the corresponding protein would probably not localize to the mitochondrion. This is similar to the situation of the previously described isoleucyl tRNA synthetase (IleRS) gene where alternative trans-splicing leads to two different size transcripts that are translated into proteins with a longer and shorter N-terminus localized to the mitochondrion and the cytosol, respectively [32].

Although we have no additional experimental verification the mass spectrometry data together with the alternative splicing profile and our previous results for the tRNA synthetase suggest that from the transcripts described above two or more proteins differing in the N-terminus are potentially translated. These findings support the hypothesis that alternative trans-splicing is a significant mechanism for the diversification of the information encoded in the *T. brucei* genome.

### MitoCarta

Corresponding to the overall coverage we detected 293 proteins predicted to be part of the mitochondrial organelle that is estimated to harbour about 1000 proteins [20,33]. For 166 of these proteins the abundance-changes between LS/SS/PC forms of the parasite were quantified (Additional file 11: Table S10). The major trend was an increase in mitochondrial protein abundance in the SS and PC cells, relative to the LS bloodstream form. The majority of changes occurred during the transition from the proliferative LS to the quiescent SS stage of the parasite, with little increase upon transition to the insect-form (Figure 3B and Additional file 1: Figure S3). A more detailed analysis of the respiratory chain confirmed this trend with the exception of the proteins from complex III and IV, which showed the main increase during the transition to the PC form (Figure 6 and Additional file 1: Figure S4) while one of the well studied bloodstream-specific proteins in the mitochondrion, the terminal alternative oxidase (TAO), decreased in abundance only after the transition to the insect-form of the parasite (mass spec ratios: SS/LS 1.6;

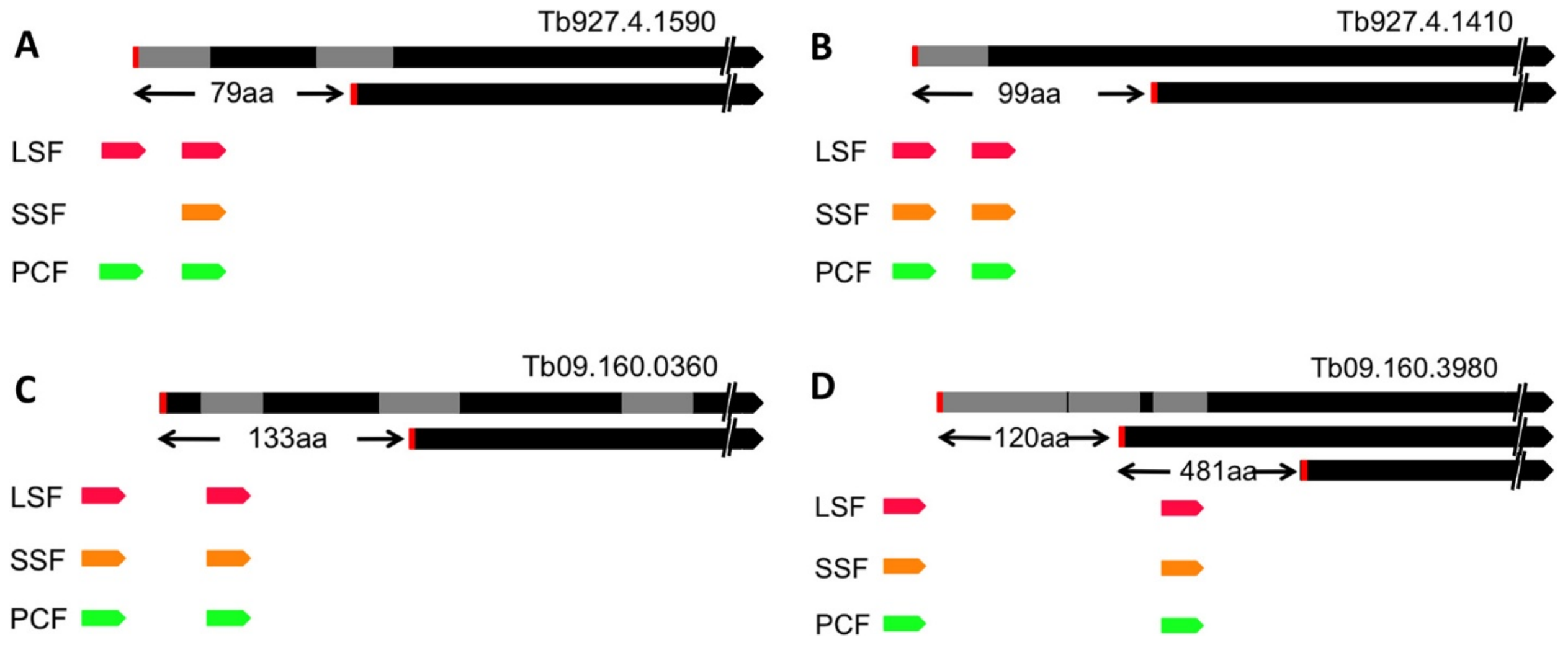

**Figure 5 Alternatively spliced transcripts and their corresponding protein products.** (**A**) Tb927.4.1590, Hypothetical protein. (**B**) Tb927.4.1410, Hypothetical protein. (**C**) Tb09.160.0360, Hypothetical protein. (**D**) Tb09.160.3980, eukaryotic translation initiation factor 4 gamma. Arrows in red, orange and green indicate splice sites detected in long slender form (LS), short stumpy form (SS) and procyclic form (PC) respectively. Black bars indicate ORFs and grey bars indicate peptide coverage. Potential start codons are shown in red.

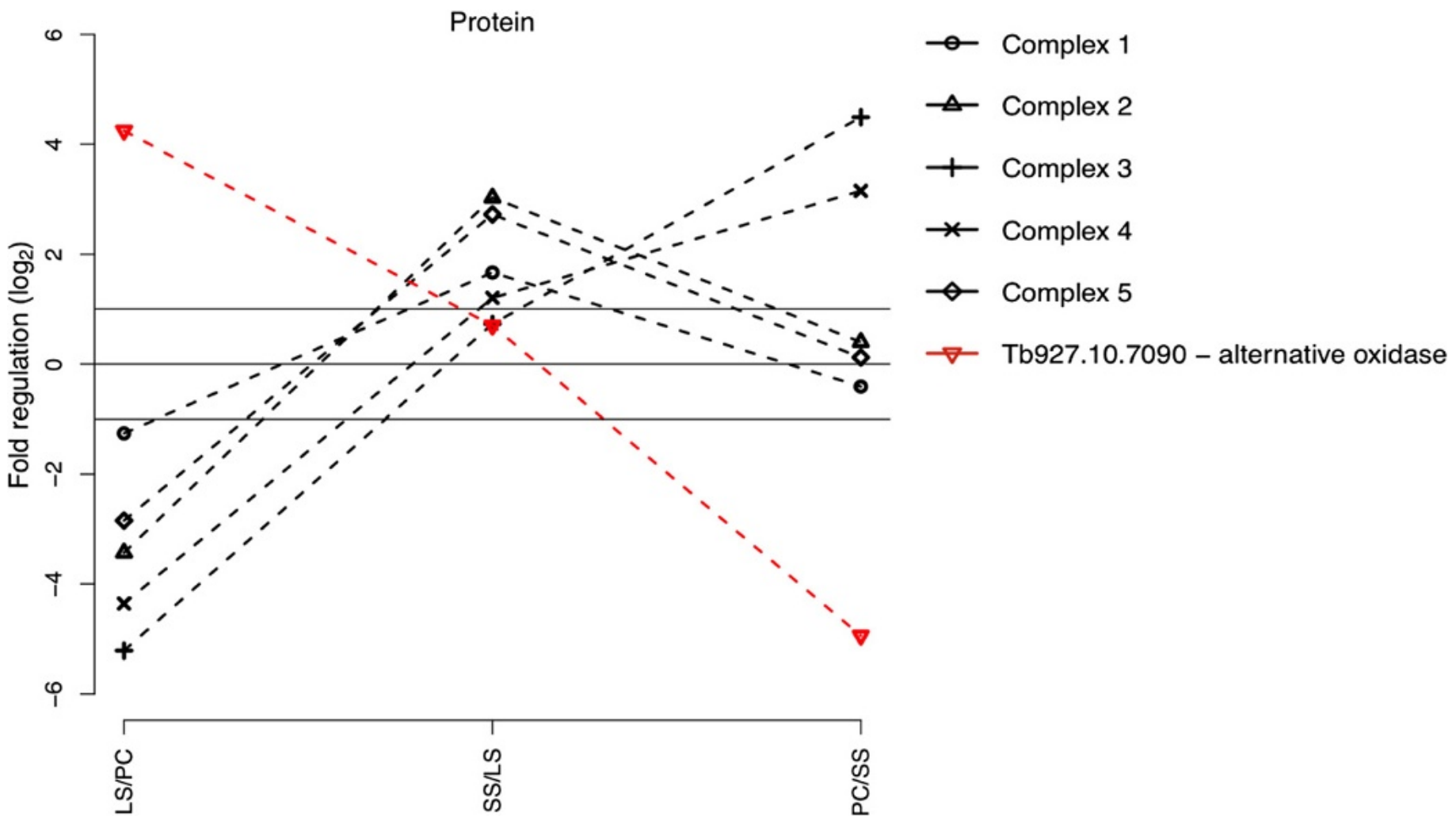

**Figure 6 Protein expression profile of the oxidative phosphorilation complexes I-V during development.** X-axis indicates the fold regulation ($log_2$) of the complexes based on the mean fold change of the individual components.

PC/SS 0.03; Tb927.10.7090). Additionally many enzymes of the procyclic-specific metabolism like the enzymes of the TCA "cycle" and the pathway converting pyruvate via acetyl-CoA to acetate are present in the SS cells at levels comparable to the insect-form. The latter pathway includes the pyruvate dehydrogenase complex subunits (PDH), the acetate:succinate CoA transferase (ASCT) and the succinyl CoA synthase (SCoAS; Additional file 1: Figure S5). In summary, we find a majority of the mitochondrial proteins detected in this study to be increased in abundance in the quiescent SS form when compared to the proliferative LS.

### Novel transcripts expressed

Kolev and co-workers recently showed the expression of 1011 previously not annotated genes in the *T. brucei* genome at the RNA level [7]. We detected peptides for 31 or 3% of these 1011 genes (Additional file 12: Table S11), while we could show the expression of about 30% of the remaining genes in the *T. brucei* genome. Four of the 31 detected proteins encoded by novel transcripts belong to the retrotransposon hot spot protein family, while the remaining, including a 41 amino acid peptide, are proteins of unknown function with no significant similarities to other proteins outside the group of trypanosomes.

### Comparison RNA/protein

Using previously published RNA expression profiling data from the same cell line (Antat1.1) and life stages we compared the relative changes of RNA and protein during the life cycle [8]. For ~40% of the genes analyzed (1529 genes, Additional file 13: Table S12) RNA and protein abundance-changes in PC/LS differed less than two fold with a Spearman's Rank correlation coefficient of rho=0.83, (Figure 7; $p<10^{-16}$). The number of genes correlating at the level of RNA and protein increased to 67% when the difference was allowed to be up to four fold (Spearman's rank correlation of rho=0.61; $p<10^{-16}$). Conversely this implied that depending on parameters and developmental stage 30% of the genes showed little to no correlation between RNA and protein abundance-changes. However, using a principle component analysis (PCA) that allows us to reduce the dimensionality/complexity of the datasets and extract common features, we could show that RNA and protein data from the individual life stages clearly cluster together (Figure 7).

## Discussion

This is the first study demonstrating the usefulness of the SILAC approach in the comparison of the total proteome of three *T. brucei* life cycle stages. We could show that fly transmissable insect-form *T. brucei* (Antat1.1; [34]) grows well in SILAC heavy isotope-adapted SDM80 medium *in vitro* with little to no Arg-$^{13}C_6$ to Pro-$^{13}C_5$ conversion detectable. This is in good agreement with the lack of the ornithine aminotransferase in the *T.brucei* genome, an enzyme essential for the conversion Arg to Pro in other systems (pers. communication F. Bringaud). Interestingly the closely related parasite Leishmania contains all enzymes necessary for the Arg to Pro conversion and thus might be less suited for a SILAC approach with labelled Arg. As expected for a gel-based approach detection was slightly biased against very small proteins, proteins with >1 trans-membrane domain and very basic proteins, the latter possibly due to trypsin based cleavage of these proteins in very small peptides that are subsequently missed in the LC mass spectrometry. However, with more than 2500 different proteins detected we achieved very good coverage of the *T. brucei* proteome (Figure 1D). The comparison with a recently published study by Urbaniak and coworkers showed overall excellent correlation of protein abundance-changes between blood (LS) and insect (PC) form cells (rho=0.83, p-value < 2.2e-16; Figure 4B) despite the fact that the growth conditions (*in vitro* vs. *in vivo* for the bloodstream cells) and the cell lines (monomorphic vs pleomorphic) were quite different. Interestingly, many mitochondrial proteins showed a decrease in fold-change between the blood and insect-form in the monomorphic when compared to the pleomorphic cell line. At this point we cannot distinguish if this is due to an increase of mitochondrial proteins in the bloodstream form, a decrease in the insect-form or a combination of both. However, it is well established that *in vitro* grown monomorphic bloodstream form cells differ in shape and motility from their *in vivo* grown pleomorphic counterparts (reviewed in [35]). Furthermore it has been speculated that the monomorphic cells are more similar to the intermediate forms of the pleomorphic cells, a differentiation step in between the LS and SS forms. When we compared our results to a recent study of the related parasite *T. congolense* we also detected strong positive correlation of protein changes during development (LS and PC), however less strong than when compared to the monomorphic *T. brucei strain* [30]. Similar to *T. brucei, the T. congolense parasite* is transmitted by the Tsetse and infects a variety of mammalian hosts. However, there are a number of differences in the mammalian host pathogenesis between the two parasites including the localization in the host [36], suggesting that there are likely differences in the underlying protein expression in the mammalian blood stage. These differences in biology could explain the discrepancies we find between the two studies, especially the significant number of proteins that are regulated in opposing directions. Additionally, the differences in the expression profiles could also be a result of the techniques applied in the two studies. Isobaric tags for relative and absolute quantitation (iTraq), which was used by Eyford and co-workers requires that the samples are individually processed, protease treated and labelled with the isobaric tags, while SILAC allowes the combined processing of the samples after growth in the heavy and light isotope media. We feel that the combined processing is an advantage of SILAC over iTraq since no variability in processing is introduced between two samples. iTraq on the other hand offers the advantage to measure up to eight samples together and thus save mass spectrometry resources and costs (for a review on iTraq see [37]).

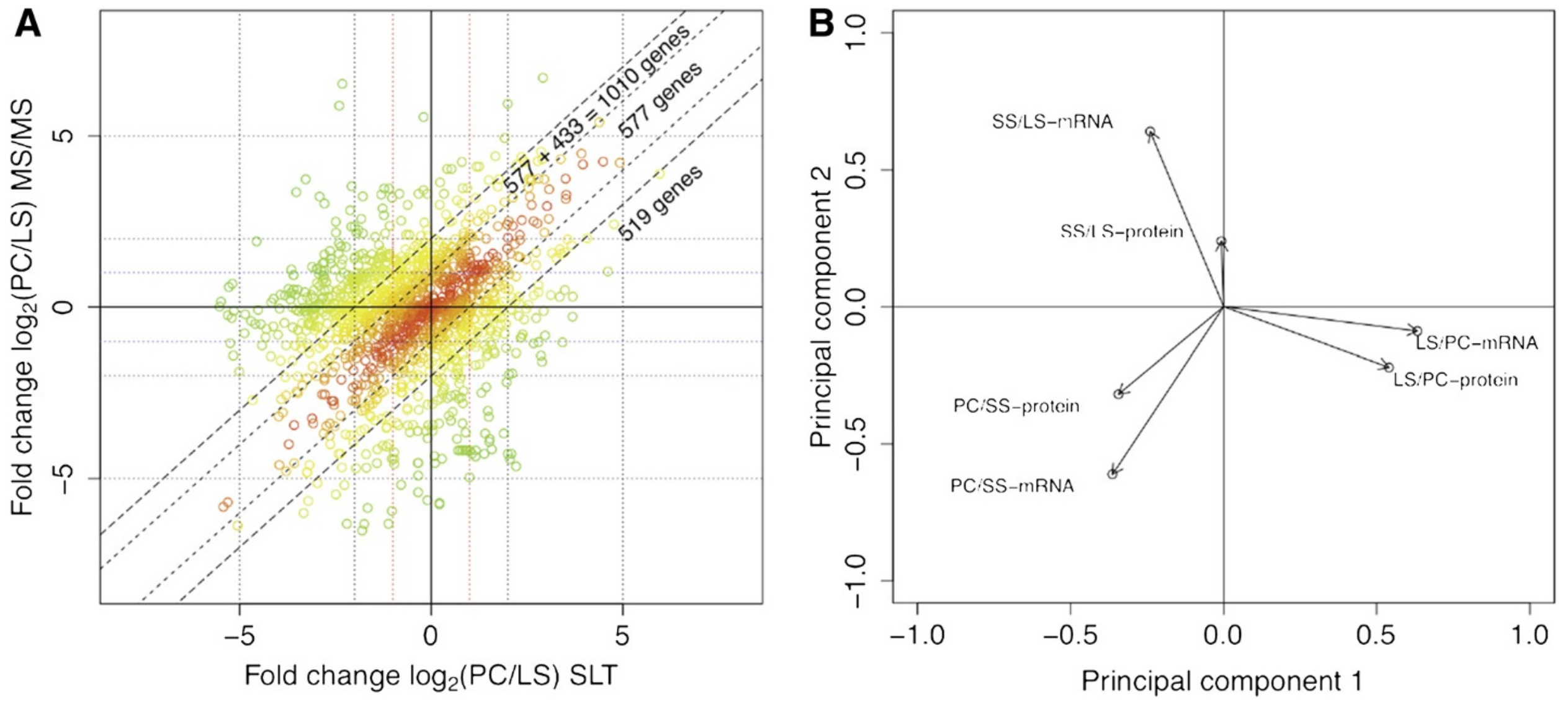

**Figure 7 Correlation of mRNA and protein changes between LS and PC *T. brucei*.** (**A**) Scatter plot comparing RNA (x-axis) and protein (y-axis) change in $\log_2$ between LS/PC. The colour of the circles change from red to green based on the correlation. The dashed lines indicate difference in fold-change ≤ 4 fold between RNA and protein. Spearman's rank correlation rho=0.61 (p-value< $2.2E^{-16}$) and Pearson's product moment correlation is 0.74 (p-value < $2.2E^{-16}$). The dotted lines indicate difference in fold-change ≤ 2 fold between RNA and protein. Spearman's rank correlation rho=0.8339 (p-value< $2.2E^{-16}$) and Pearson's product moment correlation is 0.9 (p-value < $2.2E^{-16}$). (**B**) Correlation and variability of mRNA and protein abundance changes during development based on the first two Principal components (PC).

Analyzing the changes during development we could show that about 45% of the proteins change in abundance in each of the comparisons, while more than 50% remained unchanged (Figure 1D, Figure 3A). Among the unchanged protein groups we detected 102 ribosomal proteins with mean and median fold-change of close to one in each of the life-stage comparisons (Figure 3B and Additional file 6: Table S5). This is not unexpected since ribosomal proteins are part of the cell's core machinery and as such unlikely to undergo large changes in abundance. Among the proteins that change abundance during differentiation we found very good correlation to previously published data of a number of proteins including the alternative terminal oxidase (TAO; Tb927.10.7090; [38]) a glycosomally targeted phosphatase (TbPIP39; Tb09.160.4460, [3]) and the surface transporter family proteins associated with differentiation (PAD1; Tb927.7.5930 and PAD2; Tb927.7.5940, [2]). The overall increase in protein abundance during the transition from the LS to the SS cells is explained by the specific requirements for this intermediate life cycle stage. The SS cell is adapted to the nutrients, pH and temperature of the mammalian host, while at the same time it is pre-adapted to the conditions in the tsetse midgut as particularly through the increase in abundance of mitochondrial proteins [39]. It is well established that the bloodstream form trypanosomes are devoid of cytochromes and consequently lack oxidative phosphorylation activity [6]. The data presented here now suggests that one of the final activation steps of energy production in the mitochondrion rests in the production and assembly of complexes III and IV of the oxidative phosphorylation chain, while complexes I, II and V are already present in the quiescent SS cells. This is in good agreement with the abundance of the trypanosome alternative oxidase (TAO) in the bloodstream form parasites. TAO is the terminal electron acceptor in the bloodstream cells necessary to re-oxidize the reduction equivalents produced during Glycolysis, essentially substituting complex IV. Only after the transition to the procyclic form and the activation of the regular oxidative phosphorylation chain including complex III and IV the abundance of the TAO protein decreased. At first glance the presence of many of the procyclic-specific enzymes in the SS stage is seemingly at odds with some of the metabolome data. The major end product of glucose metabolism in the SS cells, for example, has been shown to be pyruvate and not acetate as it is found in the PC cells [40]. This poses the question as to why the excess pyruvate is not converted to acetate if the enzymes of the corresponding pathway (PDH, ASCT, SCoAS) are present. We identified a potential pyruvate dehydrogenase phosphatase (Tb927.5.1660), a homolog of the PDH activating enzyme in other systems, which is upregulated at the protein level only upon transition to the insect-form. The developmental regulation of the phosphatase thus would explain the lack of ASCT activity due to a lack of acetyl CoA substrate (Additional file 1: Figure S5). Clearly this is only a hypothesis and requires experimental verification, but it demonstrates the power of the SILAC comparative approach to study the specific biology of trypanosomes. Furthermore we would like to remind the reader that the *in vitro* grown insect-form cells, although fly transmissible, are possibly different from the trypanosomes in the *in vivo* situation in the tsetse.

The recent discovery of a large number of novel transcripts prompted us to specifically search for corresponding protein products, however we only detected peptides for 3% of these transcripts. Many of the novel transcripts contain very small open reading frames (<25aa) and thus there is the possibility that these small proteins were missed on SDS PAGE, although we detected peptides belonging to a very small protein (41 aa) encoded by a novel transcript. Overall we think that a majority of the novel transcripts, although poly adenylated and capped are likely non-coding and thus present a large repertoire of potentially regulatory RNA species in the *T. brucei* genome as has recently been demonstrated for two of the novel transcripts by Michaeli and co-workers [41].

One of the enigmas of gene expression regulation is the apparent lack of correlation between the changes at the mRNA level and the protein level in some systems. A recent study, for example, described the concordance of mouse proteome and transcriptome from different mouse strains to be at a modest level ($rho=0.27$; $p<0.05$) with about 40% of the detected proteins not correlating to the changes in mRNA level [42]. On the other hand, a study in yeast showed that correlation of protein and mRNA levels following a perturbation in osmolarity was strong for the genes up-regulated at the mRNA level, however there was no correlation for mRNAs down-regulated during this process. The authors propose a model in which decrease in mRNA abundance serves purposes different from the down regulation of protein abundance during osmolarity stress [43]. In our study the comparison of LS/PC showed a similar behaviour. RNA/protein abundance-changes correlated much better when RNA was increased at least two fold (Spearman, up $r=0.36$; $p<0.0001$) than if RNA was decreased at least two fold (down, $r=0.11$; $p<0.02$), however this phenomenon was reversed when PC was compared to the SS form (up $r=0.07$, $p<0.1$; down $r=0.24$, $p<0.001$). In order to test if the lack of correlation between RNA and protein is due to biological variability at the RNA level we restricted the set to transcripts with good correlation between a previous microarray study of bloodstream and procyclic *T. brucei* and our SLT data and compared it to the protein data (Additional file 1: Figure S6). Of the 3451 genes with a robust regulation pattern at the RNA level 677 were represented in our mass spectrometry data. The correlation coefficient of the restricted dataset ($rho=0.35$; $p<2.2e^{-16}$) was only marginally higher than the entire dataset ($rho=0.28$; $p<2.2e^{-16}$) thus even for transcripts with a robust expression pattern, protein abundance-changes do not correlate very well. Although we cannot entirely exclude that the difference in regulation at the RNA and protein level is due to the variability in biological replicates it seems likely that a significant portion of the protein abundance-changes are due to regulation of translation or protein stability and thus are not reflected at the RNA level.

## Conclusions

In summary we show (i) the effective use of SILAC for the analysis of whole proteome changes during the life cycle in *T. brucei*, especially also to uncover potential regulatory proteins that control the differentiation (ii) the likely N-termini of 109 proteins and that number of them possibly produce several N-terminal isoforms; (iii) a surprising difference between the protein and RNA changes during the life cycle, that are possibly explained by a significant level of translational/posttranslational regulation that has yet to be explored, (iv) that the massive changes, especially in the mitochondrial proteome of the quiescent SS form support its status as a distinctly differentiated subpopulation of cells committed to life in the fly.

## Methods

### Cells and growth conditions

Procyclic Antat 1.1 [44] cells were pre-adapted to SDM 80 with 10% dialyzed fetal bovine serum (FCS; 10,000 molecular weight cut-off; Amimed) for seven days at 27°C; cells were then washed with PBS and transferred into 10 ml SDM 80 with 10% dialyzed fetal bovine serum (FCS) and the stable isotope labelled amino acids Arginine (1,1 mM; $^{13}C_6$, 99%, CIL) and Lysine (0,4 mM; $^{13}C_6$, 99%, CIL) that replaced the non labelled amino acids. Cells were harvested after 0, 2 and 11 cell division cycles (doubling time 8 hours) and the incorporation of stable isotope labelled amino acids was monitored by mass spectrometry. For the final experiments cells were harvested $5 \times 10^7$ cells ($5 \times 10^6$ cells/ml) after 11 cell division cycles in SDM with stable isotope labelled amino acids. Cells were washed with PBS and lysed directly in Laemmli buffer at a concentration of $5{*}10^5$ cells/µl. Lysates were stored at −80°C.

Bloodstream form cells were grown in rats (Wistar). Long slender cells were harvested at day three post-intraperitoneal injection of $5 \times 10^6$ cells at a density of $5 \times 10^7$ cells/ml blood. For short stumpy cells rats were immunosuppressed with 200 mg/kg cyclophosphoamide (Sigma) 24 hours prior to intraperitoneal injection of $1 \times 10^6$ cells. Short stumpy cells were harvested at day five from immunosuppressed rats at a density of $5\text{-}8 \times 10^8$ cells/ml blood. 85% of cells harvested at day five displayed the typical short stumpy appearance with the flagellum significantly shortened, while >90% of cells harvested at day three showed a long slender appearance with the flagellum extending anterior to the cell body, as seen by light microscopy after methanol fixation. Bloodstream form cells were purified from whole rat blood using DE-52 anion exchange resin (Whatman) equilibrated to pH 8 with a bicine glucose buffer (50 mM bicine, 50 mM NaCl, 5 mM KCl, 50 mM glucose).

### Mass spectrometry

Two technical replicates were done each using $5 \times 10^6$ procyclic cells that were mixed with $5 \times 10^6$ LS or SS cells in Laemmli buffer, boiled at 95°C for 5 min and resolved on a 10% acryl amide gel (1,0 mm). Gels were stained with standard Coomassie Brilliant Blue G-250 solution (2.5 g Coomassie Brilliant Blue G-250 from Applichem in 450 ml methanol, 100 ml acetic acid and 400 ml water) and destained in the same solution without Coomassie. Subsequently gels were stored in 20% (v/v) ethanol at 4°C until MS analysis that was done in duplicate on two SDSPAGE lanes (within 2 days). For this, each gel lane was cut into ten bands. Each band was cut into several little cubes that were transferred to a low-binding reagent tube (Sarstedt, Nümbrecht, Germany). Gel slices were washed with 50 mM Tris/HCl pH 8 (Tris buffer) and Tris buffer/acetonitrile (LC-MS grade, Fluka, Buchs, Switzerland) 50/50 before protein reduction with 50 mM DTT (Fluka, Buchs, Switzerland) in Tris buffer for 30 min at 37°C, and alkylation with 50 mM iodoacetamide (Fluka, Buchs, Switzerland) in Tris buffer for 30 min at 37°C in the dark. After washing with Tris buffer and dehydration with acetonitrile the gel cubes were soaked with trypsin solution composed of 10 ng/ml trypsin (Promega) in 20 mM Tris/HCl pH 8, 0.01% ProteaseMax (Promega) for 30 min on ice. Gel cubes were covered by addition of 5–10 ml 20 mM Tris/HCl before digestion for 60 min at 50°C. The supernatant liquid was combined with a single gel extract performed with 20 ml 20% (v/v) formic acid (Merck) in polypropylene HPLC vials. An aliquot of 10 ml from each digest was loaded onto a self-made pre-column (Magic C18, 5 mm, 300 Å, 0.15 mm i.d. x 30 mm length) at a flow rate of ~5 ml/min with solvent A (0.1% formic acid in water/acetonitrile 98:2). After loading, peptides were eluted in back flush mode onto the analytical nano-column (Magic C18, 5 mm, 100 Å, 0.075 mm i.d. × 75 mm length) using an acetonitrile gradient of 5% to 40% solvent B (0.1% formic acid in water/acetonitrile 4.9:95) in 60 min at a flow rate of ~400 nl/min. The column effluent was directly coupled to an LTQ-orbitrap XL mass spectrometer (ThermoFisher, Bremen, Germany) via a nanospray ESI source operated at 1700 V. Data acquisition was made in data dependent mode with precursor ion scans recorded in the Fourier transform detector (FT) with resolution of 60’000 (@ m/z =400) parallel to five fragment spectra of the most intense precursor ions in the linear iontrap. CID mode settings were: Wideband activation on; precursor ion selection between m/z range 360–1400;

intensity threshold at 500; precursors excluded for 15 sec. Further tune parameter settings were: Max. injection time LTQ MS2 = 200 ms, orbitrap MS 500 ms; automatic gain control orbitrap = $5 \times 10^5$, LTQ MS = $3 \times 10^4$, MS2 = $1 \times 10^4$.

Mass Spectrometry data and SILAC ratio interpretation was made with MaxQuant version 1.1.1.36 run under Windows7 against a *T. brucei* sequence database (version available in May 2011) from the Wellcome Trust Sanger Institute Pathogen Sequencing Unit. The default contamination database in Andromeda (MaxQuant) was searched together with the target database. We applied the following MaxQuant default settings: for precursor masses in the first search (+/− 20ppm), in the second search (re-calibrated mass values; +/− 6ppm). For fragment spectra default was set to +/− 0.5 Da and only the top 6 peaks per mass interval of 100 were kept (peak filtering; [45]). Other Maxquant parameters were: Peptide and protein FDR set at 1%; carbamidomethylcystein set as fixed modification; allowed dynamic modifications were Met oxidation, protein N-terminal acetylation, Phosphorylation on Ser/Thr/Tyr. For SILAC ratio at least two unique or razor peptides without modification were required. A strict trypsin cleavage rule was applied i.e. cleavage c-terminal after K or R, no cleavage if a P follows R or K!. The normalisation was done using the default settings in MaxQuant [46].

### Western blotting

For western blotting total cell protein from $1 \times 10^7$ cells of bloodstream (long slender, short stumpy) and procyclic cells was resolved on 10% SDSPAGE, transferred to a PVDF membrane (Immobilon FL, Millipore) and probed with antibodies shown in Additional file 3: Table S1. Signals were quantified using the Odyssey Infrared Imaging System (LI-COR).

### Bioinformatics

#### *Analysing MSMS data*

Perl (perl, v5.10.1), unix (in Ubuntu 10.04.3 LTS (Lucid Lynx)) and R (R version 2.14.1 (2011-12-22)) scripts were used to analyse RNA and protein data. The statistical analyses were performed using R scripts. The scripts will be provided upon request. All plots except the circular visualisation of protein abundance along the chromosomes of *T. brucei*, which was produced using Circos software [47] were produced using R and supportive packages.

#### *Detecting novel transcripts and genes with N-terminal extensions*

Peptide coverage for novel transcripts and n-terminal extensions was detected using Trans Proteomic Pipeline (TPP, [48]). The raw files containing MS/MS spectra were converted into mzXML files and subsequently files were used for database searches. The database searches were done using the X!Tandem search-engine. Protein databases were created based on DNA sequences of the novel transcripts and genes with N-terminal extensions for the database search. Precursor mass tolerance was set to −2 and +4 Da. We used this mass tolerance window together with the other default parameters for the X!Tandem database search. The output files of X!Tandem search engine were converted into pepXML (XML format) files, which were used for subsequent peptide level analysis. Validation of peptide spectrum assignment was done using PeptideProphet with the "accurate mass binning" option enabled. The minimum peptide length considered in the analysis is seven amino acids. The results were filtered to a 1% FDR at the peptide and protein levels. Finally, the protein identifications were converted into protXML files, which contain the information for protein level analysis.

#### *Principal component analysis*

The SLT and MS/MS data sets are modelled as a numerical matrix of $M \times N$, where $M$ is number of genes under $N$ comparisons (3 × SLT and 3 × MS/MS datasets). The eigenvalues and eigenvectors were computed by solving the following equation:

$$\left(\bar{\bar{C}} - \lambda\right)\bar{V} = 0$$

Where $\bar{\bar{C}}$ is the correlation matrix, i.e. square symmetric and $\lambda$ is the eigenvalue and $\bar{V}$ is the eigenvector. The size of the eigenvectors is equal to number of variables and the eigenvectors are orthogonal to each other, i.e. $\bar{V}_1 \bullet \bar{V}_2 = 0$.

We have used R, rgl and lattice packages to compute the matrix of variable loadings, rotated data and to plot rotated variables.

The data can be accessed online through: http://splicer.unibe.ch/, TritrypDB ( http://tritrypdb.org) and Proteome Commons (http://www.proteomecommons.org).

## Additional files

**Additional file 1: Figure S1.** Comparison of procyclic *T. brucei* cell growth in different media. (**A**) Cells grown in SDM 79/80 with regular FCS and SDM 80 with dialyzed FCS (dFCS). (**B**) Comparison of procyclic *T. brucei* cell growth of cells in SDM 80 dFCS regular amino acids and cells in SDM 80, dFCS and heavy isotope labeled amino acids (Arg; Lys). **Figure S2.** Labelling efficiency of procyclic *T. brucei* whole cell proteome after zero, two and 11 cell division cycles. (**A**) Box plot depicting mean ratios of heavy/light peptides from the whole proteome. (**B**) Shows the relative abundance shift of mass spectrometry peaks (m/z) during the course of labelling (0, 2 and 11 cell division cycles) with heavy amino acids towards higher m/z ratios. Data corresponds to one peptide of the triosephosphate isomerase gene (Tb11.02.3210). **Figure S3.** Expression profile of 166 mitochondrial proteins during *T. brucei* development. Each circle indicates a cluster of co-regulated proteins. Circle size is proportional to the number of proteins in the cluster. The midpoint of the circle marks the mean fold change between the life cycle stages. Expression profile is depicted as fold change ($log_2$) between the different life cycle stages. **Figure S4.** Protein expression profile of oxidative phosphorylation complexes I-V during development. (**A and B**) Complex one. (**C and D**) Complex two. (**E and F**) Complex three. (**G and H**) Complex four. (**I and J**) Complex five. Each gene is assigned a sign. x-axis indicate the fold regulation in $log_2$ between life stages. **Figure S5.** Procyclic specific metabolism genes and their regulation during development. (**A**) Abundance profile of eight different proteins likely involved in pyruvate to acetate conversion in the mitochondrion of insect form trypanosomes. (**B**) Model depicting the conversion of pyruvate to acetate. Green arrow indicates activation of the pyruvate dehydrogenase complex through dephosphorylation. **Figure S6.** Gene expression comparison between SLT, microarray and MS/MS. (**A**) A set of highly correlated genes between microarray (Jensen *et al.*) and SLT (Nilsson *et al.*). The Pearson's product moment correlation is 0.82 (p-value < 2.2e-16). (**B**) Correlation of the same set between SLT and MS/MS. The Pearson's product moment correlation is 0.35 (p-value < 2.2e-16). (**C**) Correlation of mRNA and protein changes between LS and PC. The Pearson's product moment correlation is 0.28 (p-value < 2.2e-16). Colors of the circles in the scatter plot indicate correlation between the two studies (red high correlation, green low correlation).

**Additional file 2: Table S13.** Peptide information of PC/LS and PC/SS.

**Additional file 3: Table S1.** Antibodies and conditions used in this study.

**Additional file 4: Table S2.** Proteins detected in both the heavy fraction PC and light fraction LS. Raw – The raw excel sheet generated by MaxQuant system on heavy fraction PC and light fraction LS.
**Table S3** – Proteins detected in both the heavy fraction PC and light fraction SS. **Raw** – The raw excel sheet generated by MaxQuant system on heavy fraction PC and light fraction SS.

**Additional file 5: Table S4.** Proteins detected in both PC/LS and PC/SS experiments (overlaps).

**Additional file 6: Table S5.** GO terms with gene IDs.

**Additional file 7: Table S6.** Proteins only detected in PC/LS experiment.

**Additional file 8: Table S7.** Proteins only detected in PC/SS experiment.

**Additional file 9: Table S8.** Proteins having difference in fold change >= 5 fold between Urbaniak and our study.

**Additional file 10: Table S9.** Alternative splicing to diversify protein production.

**Additional file 11: Table S10.** Change of mitochondrial protein abundance during differentiation.

**Additional file 12: Table S11.** Coverage of novel transcripts.

**Additional file 13: Table S12.** Overlaps of SLT and MS/MS data.

### Competing interests
The authors declare: no financial relationships with any organisations that might have an interest in the submitted work in the previous 5 years; no other relationships or activities that could appear to have influenced the submitted work.

### Authors' contributions
TO and MH conceived and designed the experiments. DW, SB and TO performed the experiments. KG, DW and MH performed the analysis. KG and TO wrote the paper. All authors read and approved the final manuscript.

### Acknowledgements
This research was funded by the Swiss National Science Foundation (31003A-125194). We thank A. Schneider, I. Roditi, C. Clayton and J. Lukes for antibodies.

### Author details
[1]Institute for Cell Biology, University of Bern, Bern, Switzerland. [2]Department of Clinical Research, University of Bern, Bern, Switzerland. [3]Graduate School for Cellular and Biomedical Sciences, University of Bern, Bern, Switzerland.

### References
1. Vassella E, Reuner B, Yutzy B, Boshart M: **Differentiation of African trypanosomes is controlled by a density sensing mechanism which signals cell cycle arrest via the cAMP pathway.** *J Cell Sci* 1997, **110**(Pt 21):2661–2671.
2. Dean S, Marchetti R, Kirk K, Matthews KR: **A surface transporter family conveys the trypanosome differentiation signal.** *Nature* 2009, **459**(7244):213–217.
3. Szoor B, Ruberto I, Burchmore R, Matthews KR: **A novel phosphatase cascade regulates differentiation in Trypanosoma brucei via a glycosomal signaling pathway.** *Genes Dev* 2010, **24**(12):1306–1316.
4. Opperdoes FR, Szikora JP: **In silico prediction of the glycosomal enzymes of Leishmania major and trypanosomes.** *Mol Biochem Parasitol* 2006, **147**(2):193–206.
5. Parsons M: **Glycosomes: parasites and the divergence of peroxisomal purpose.** *Mol Microbiol* 2004, **53**(3):717–724.
6. Priest JW, Hajduk SL: **Developmental regulation of mitochondrial biogenesis in Trypanosoma brucei.** *J Bioenerg Biomembr* 1994, **26**(2):179–191.
7. Kolev NG, Franklin JB, Carmi S, Shi H, Michaeli S, Tschudi C: **The transcriptome of the human pathogen Trypanosoma brucei at single-nucleotide resolution.** *PLoS Pathog* 2010, **6**(9):e1001090.
8. Nilsson D, Gunasekera K, Mani J, Osteras M, Farinelli L, Baerlocher L, Roditi I, Ochsenreiter T: **Spliced leader trapping reveals widespread alternative splicing patterns in the highly dynamic transcriptome of Trypanosoma brucei.** *PLoS Pathog* 2010, **6**(8):e1001037.
9. Siegel TN, Hekstra DR, Wang X, Dewell S, Cross GA: **Genome-wide analysis of mRNA abundance in two life-cycle stages of Trypanosoma brucei and identification of splicing and polyadenylation sites.** *Nucleic Acids Res* 2010, **38**(15):4946–4957.
10. Jensen BC, Sivam D, Kifer CT, Myler PJ, Parsons M: **Widespread variation in transcript abundance within and across developmental stages of Trypanosoma brucei.** *BMC Genomics* 2009, **10**(1):482.
11. Kabani S, Fenn K, Ross A, Ivens A, Smith TK, Ghazal P, Matthews K: **Genome-wide expression profiling of in vivo-derived bloodstream parasite stages and dynamic analysis of mRNA alterations during synchronous differentiation in Trypanosoma brucei.** *BMC Genomics* 2009, **10**(1):427.
12. Koumandou VL, Natesan SK, Sergeenko T, Field MC: **The trypanosome transcriptome is remodelled during differentiation but displays limited responsiveness within life stages.** *BMC Genomics* 2008, **9**:298.
13. Webb H, Burns R, Ellis L, Kimblin N, Carrington M: **Developmentally regulated instability of the GPI-PLC mRNA is dependent on a short-lived protein factor.** *Nucleic Acids Res* 2005, **33**(5):1503–1512.
14. Berberof M, Vanhamme L, Tebabi P, Pays A, Jefferies D, Welburn S, Pays E: **The 3'-terminal region of the mRNAs for VSG and procyclin can confer stage specificity to gene expression in Trypanosoma brucei.** *EMBO J* 1995, **14**(12):2925–2934.
15. Furger A, Schurch N, Kurath U, Roditi I: **Elements in the 3' untranslated region of procyclin mRNA regulate expression in insect forms of Trypanosoma brucei by modulating RNA stability and translation.** *Mol Cell Biol* 1997, **17**(8):4372–4380.
16. Hehl A, Vassella E, Braun R, Roditi I: **A conserved stem-loop structure in the 3' untranslated region of procyclin mRNAs regulates expression in Trypanosoma brucei.** *Proc Natl Acad Sci U S A* 1994, **91**(1):370–374.
17. Saas J, Ziegelbauer K, von Haeseler A, Fast B, Boshart M: **A developmentally regulated aconitase related to iron-regulatory protein-1 is localized in the cytoplasm and in the mitochondrion of Trypanosoma brucei.** *J Biol Chem* 2000, **275**(4):2745–2755.
18. Priest JW, Hajduk SL: **Developmental regulation of Trypanosoma brucei cytochrome c reductase during bloodstream to procyclic differentiation.** *Mol Biochem Parasitol* 1994, **65**(2):291–304.
19. Mayho M, Fenn K, Craddy P, Crosthwaite S, Matthews K: **Post-transcriptional control of nuclear-encoded cytochrome oxidase subunits in Trypanosoma brucei: evidence for genome-wide conservation of life-** cycle stage-specific regulatory elements. *Nucleic Acids Res* 2006, **34**(18):5312–5324.
20. Panigrahi AK, Ogata Y, Zikova A, Anupama A, Dalley RA, Acestor N, Myler PJ, Stuart KD: **A comprehensive analysis of Trypanosoma brucei mitochondrial proteome.** *Proteomics* 2009, **9**(2):434–450.
21. Bridges DJ, Pitt AR, Hanrahan O, Brennan K, Voorheis HP, Herzyk P, de Koning HP, Burchmore RJ: **Characterisation of the plasma membrane subproteome of bloodstream form Trypanosoma brucei.** *Proteomics* 2008, **8**(1):83–99.
22. Oberholzer M, Langousis G, Nguyen HT, Saada EA, Shimogawa MM, Jonsson ZO, Nguyen SM, Wohlschlegel JA, Hill KL: **Independent analysis of the flagellum surface and matrix proteomes provides insight into flagellum signaling in mammalian-infectious Trypanosoma brucei.** *Mol Cell Proteomics* 2011, **10**(10):M111 010538.
23. Broadhead R, Dawe HR, Farr H, Griffiths S, Hart SR, Portman N, Shaw MK, Ginger ML, Gaskell SJ, McKean PG, *et al*: **Flagellar motility is required for the viability of the bloodstream trypanosome.** *Nature* 2006, **440**(7081):224–227.
24. Vertommen D, Van Roy J, Szikora JP, Rider MH, Michels PA, Opperdoes FR: **Differential expression of glycosomal and mitochondrial proteins in the two major life-cycle stages of Trypanosoma brucei.** *Mol Biochem Parasitol* 2008, **158**(2):189–201.
25. Colasante C, Ellis M, Ruppert T, Voncken F: **Comparative proteomics of glycosomes from bloodstream form and procyclic culture form Trypanosoma brucei brucei.** *Proteomics* 2006, **6**(11):3275–3293.
26. Gruhler A, Schulze WX, Matthiesen R, Mann M, Jensen ON: **Stable isotope labeling of Arabidopsis thaliana cells and quantitative proteomics by mass spectrometry.** *Mol Cell Proteomics* 2005, **4**(11):1697–1709.
27. Walther DM, Mann M: **Accurate quantification of more than 4000 mouse tissue proteins reveals minimal proteome changes during aging.** *Mol Cell Proteomics* 2010, **10**(2):M110 004523.
28. de Godoy LM, Olsen JV, Cox J, Nielsen ML, Hubner NC, Frohlich F, Walther TC, Mann M: **Comprehensive mass-spectrometry-based proteome quantification of haploid versus diploid yeast.** *Nature* 2008, **455**(7217):1251–1254.
29. Lanham SM, Godfrey DG: **Isolation of salivarian trypanosomes from man and other mammals using DEAE-cellulose.** *Exp Parasitol* 1970, **28**(3):521–534.
30. Urbaniak MD, Guther ML, Ferguson MA: **Comparative SILAC proteomic analysis of Trypanosoma brucei bloodstream and procyclic lifecycle stages.** *PLoS One* 2012, **7**(5):e36619.
31. Eyford BA, Sakurai T, Smith D, Loveless B, Hertz-Fowler C, Donelson JE, Inoue N, Pearson TW: **Differential protein expression throughout the life cycle of Trypanosoma congolense, a major parasite of cattle in Africa.** *Mol Biochem Parasitol* 2011, **177**(2):116–125.
32. Rettig J, Wang Y, Schneider A, Ochsenreiter T: **Dual targeting of Isoleucyl tRNA synthetase in Trypanosoma brucei is mediated through alternative trans-splicing.** *Nucleic Acids Res* 2011.
33. Zhang X, Cui J, Nilsson D, Gunasekera K, Chanfon A, Song X, Wang H, Xu Y, Ochsenreiter T: **The Trypanosoma brucei MitoCarta and its regulation and splicing pattern during development.** *Nucleic Acids Res* 2010, **38**(21):7378–7387.
34. Vassella E, Oberle M, Urwyler S, Renggli CK, Studer E, Hemphill A, Fragoso C, Butikofer P, Brun R, Roditi I: **Major surface glycoproteins of insect forms of Trypanosoma brucei are not essential for cyclical transmission by tsetse.** *PLoS One* 2009, **4**(2):e4493.
35. Matthews KR, Ellis JR, Paterou A: **Molecular regulation of the life cycle of African trypanosomes.** *Trends Parasitol* 2004, **20**(1):40–47.
36. Luckins AG, Gray AR: **An extravascular site of development of Trypanosoma congolense.** *Nature* 1978, **272**(5654):613–614.
37. Christoforou AL, Lilley KS: **Isobaric tagging approaches in quantitative proteomics: the ups and downs.** *Anal Bioanal Chem* 2012, **404**(4):1029–1037.
38. Chaudhuri M, Ajayi W, Temple S, Hill GC: **Identification and partial purification of a stage-specific 33 kDa mitochondrial protein as the alternative oxidase of the Trypanosoma brucei brucei bloodstream trypomastigotes.** *J Eukaryot Microbiol* 1995, **42**(5):467–472.
39. Tyler KM, Matthews KR, Gull K: **The bloodstream differentiation-division of Trypanosoma brucei studied using mitochondrial markers.** *Proc Biol Sci* 1997, **264**(1387):1481–1490.
40. van Grinsven KW, Van Den Abbeele J, Van den Bossche P, van Hellemond JJ, Tielens AG: **Adaptations in the glucose metabolism of procyclic Trypanosoma brucei isolates from tsetse flies and during differentiation of bloodstream forms.** *Eukaryot Cell* 2009, **8**(8):1307–1311.
41. Michaeli S, Doniger T, Gupta SK, Wurtzel O, Romano M, Visnovezky D, Sorek R, Unger R, Ullu E: **RNA-seq analysis of small RNPs in Trypanosoma brucei reveals a rich repertoire of non-coding RNAs.** *Nucleic Acids Res* 2012, **40**(3):1282–1298.
42. Ghazalpour A, Bennett B, Petyuk VA, Orozco L, Hagopian R, Mungrue IN, Farber CR, Sinsheimer J, Kang HM, Furlotte N, *et al*: **Comparative analysis of proteome and transcriptome variation in mouse.** *PLoS Genet* 2011, **7**(6):e1001393.
43. Lu P, Vogel C, Wang R, Yao X, Marcotte EM: **Absolute protein expression profiling estimates the relative contributions of transcriptional and translational regulation.** *Nat Biotechnol* 2007, **25**(1):117–124.
44. Le Ray D, Barry JD, Easton C, Vickerman K: **First tsetse fly transmission of the "AnTat" serodeme of Trypanosoma brucei.** *Ann Soc Belg Med Trop* 1977, **57**(4–5):369–381.
45. Cox J, Neuhauser N, Michalski A, Scheltema RA, Olsen JV, Mann M: **Andromeda: a peptide search engine integrated into the MaxQuant environment.** *J Proteome Res* 2011, **10**(4):1794–1805.
46. Cox J, Mann M: **MaxQuant enables high peptide identification rates, individualized p.p.b.-range mass accuracies and proteome-wide protein quantification.** *Nat Biotechnol* 2008, **26**(12):1367–1372.
47. Krzywinski M, Schein J, Birol I, Connors J, Gascoyne R, Horsman D, Jones SJ, Marra MA: **Circos: an information aesthetic for comparative genomics.** *Genome Res* 2009, **19**(9):1639–1645.
48. Keller A, Eng J, Zhang N, Li XJ, Aebersold R: **A uniform proteomics MS/MS analysis platform utilizing open XML file formats.** *Mol Syst Biol* 2005, **1**:2005 0017.